# On Possible Influence of Space Weather on Agricultural Markets: Necessary Conditions and Probable Scenarios

**L. Pustil'nik**[1, 2*] **and G. Yom Din**[1, 3**]

[1] *Tel Aviv University, Tel Aviv, 69978 Israel*
[2] *Golan Research Institute, Katzrin, 12900 Israel*
[3] *Open University of Israel, Raanana, 43537 Israel*

\* E-mail: levpust@post.tau.ac.il
\*\* E-mail: gregory@openu.ac.il

**Abstract**. We present the results of study of a possible relationship between the space weather and terrestrial markets of agricultural products. It is shown that to implement the possible effect of space weather on the terrestrial harvests and prices, a simultaneous fulfillment of three conditions is required: 1) sensitivity of local weather (cloud cover, atmospheric circulation) to the state of space weather; 2) sensitivity of the area-specific agricultural crops to the weather anomalies (belonging to the area of risk farming); 3) relative isolation of the market, making it difficult to damp the price hikes by the external food supplies. Four possible scenarios of the market response to the modulations of local terrestrial weather via the solar activity are described. The data sources and analysis methods applied to detect this relationship are characterized. We describe the behavior of 22 European markets during the medieval period, in particular, during the Maunder minimum (1650–1715). We demonstrate a reliable manifestation of the influence of space weather on prices, discovered in the statistics of intervals between the price hikes and phase price asymmetry. We show that the effects of phase price asymmetry persist even during the early modern period in the U.S. in the production of the durum wheat. Within the proposed approach, we analyze the statistics of depopulation in the eighteenth and nineteenth century Iceland, induced by the famine due to a sharp livestock reduction owing to, in its turn, the lack of foodstuff due to the local weather anomalies. A high statistical significance of temporal matching of these events with the periods of extreme solar activity is demonstrated. We discuss the possible consequences of the observed global climate change in the formation of new areas of risk farming, sensitive to space weather.

## 1. HISTORICAL INTRODUCTION

The problem of the possible influence of solar activity on the Earth's agriculture has an almost 300 year-long history. One of its first references appears in the description of the British Royal Society (the analog of the Academy of Sciences) by a famous professor of theology and, concurrently, the father of European satire Jonathan Swift in his book devoted to the third Gulliver's Travel to the island of Laputa [1]. In this keen satire, Swift, describing the main activities of the Laputans[1] mentioned the following as their two main phobias:

1) Under the influence of a heavenly body (a comet), the Earth, got caught in its glowing tail, will undergo a period of "global warming," threatening imminent death to all living creatures;

2) The Sun will be covered with its own feces (spots) and will cease sending the light and

---

[1] As Swift asserted, the Laputans devoted most of their time to the attempts of studying the matter using super-strong magnets and researching the skies with giant telescopes. We can't but note that these engagements do ever since (for 300 years by now) hold the leading positions in the scientific programs ventured by mankind.

warmth to Earth (followed by the "global cooling").

Given that the sources of anxiety of the Laputans described by Swift reflect the phobias and rumors prevalent in the "enlightened society," contemporary to Swift, we can only be surprised how stable they are, preserved to this day in the fashionable science fiction novels, disaster movies, and headlines.

The subsequent statement about the possible im-pact of space weather and solar activity on the agriculture was made 75 years later by the father of the European observational astronomy, renowned Sir William Herschel. Comparing the data on the grain prices from the fundamental work of Adam Smith [2], titled "An Inquiry into the Nature and Causes of the Wealth of Nations" [2] (1776) with the data on the number of sunspots over the same period, Herschel [3] made a far-reaching conclusion: "Five long periods of sunspot scarcity coincide with the periods of in-creased wheat prices." This work was published in the proceedings of the British Royal Society and met with hostility by the other members of the academy, who have by then already abandoned the earlier illusions of Swift. In an extremely harsh review of one of the leaders of the Royal Society, Professor Herschel was given a mocking title of the "King of Absurdity." Given the gigantic authority of Herschel, who has by then discovered a new planet Uranus (initially named the "Georgian Planet" in honor of his new patron, King George III), and his status of the Royal Astronomer, the obstruction of his work demonstrates an absolute rejection by the scientific community of the time of the idea of possible influence of solar activity on Earth.

The next, but not the last[3] victim of interest to this problem was the famous establisher of the mathematical school in political economy, one of the founders of the theory of marginal utility, Professor William Jevons. Jevons [5–7] was one of the first to point to the cyclicity in the development of the economy and, in particular, to the periodic recur-rence of economic crises with an average period between them of about 10.2 years. Discovering an exceptional closeness of the period he found between the economic crises with the period of the 11-year sunspot cycle discovered a short while before, Jevons suggested that solar activity somehow modulates the economic activity. As a possible causal chain explaining this coincidence, Jevons suggested that in the years of "unfavorable" solar activity there occur some weather anomalies, leading to crop failures and higher food prices, decreasing demand for industrial products and, ultimately, generation of the stock exchange crises. Jevons saw India and South-East Asia as the "soft spots," potentially sensitive to such anomalies. As of the weather anomalies, he considered the weakening or cessation of the summer monsoon responsible for the irrigation of this area. Extrapolating the coincidence of the five previous exchange crises with the periods of solar activity minima, and activity surges immediately after the minima, Jevons had the audacity to predict the future economic crisis in the years close to the next solar activity minimum in 1879, which naturally caused a strong reaction in the stock exchange and in press. Since the promised crisis did not occur neither in 1879, nor later[4], the theory of Jevons was completely discredited, and the concept of "sunspot equilibria" appeared in the economy[5].

---

[2] The acquaintance with which A. S. Pushkin reckoned among the advantages of his "good friend," the notorious Eugene Onegin.

[3] On this count we have to first of all recall of the name of Professor Chizhevsky [4], who was one of the first to use the nowadays popular term "space weather" and did a lot to promote the then heretical topics of solar-terrestrial relations in Russia.

The further discovery of the exceptional consistency (within 0.1%) of the solar radiation levels reaching the Earth (which has therefore obtained the name of the "solar constant"), has for a long while deprived the supporters of the influence of solar activity on the terrestrial processes of any physical arguments.

This "pessimistic" period lasted until new channels of influence of solar activity on Earth were detected, which are nowadays united by the term "space weather."

## 2. SPACE WEATHER AS A FACTOR OF INFLUENCE ON TERRESTRIAL PROCESSES

### 2.1. Introduction to Space Weather

The term "space weather" is now understood as a combination of factors of extraterrestrial origin, capable of exerting a significant influence on Earth: in particular, the solar wind and high-energy cosmic rays, solar flares and coronal mass ejections.[6] The phenomenon of solar activity, caused by the dynamo processes of cyclic generation of magnetic fields on the Sun is fundamental in this group of processes. As a result, the sunspots and coronal holes occur on the surface of the Sun with the formation of a high-temperature solar corona, then its outer part expands and turns into the solar wind. In its turn, the solar wind, blowing over the Earth's magnetosphere, forms its boundaries. The magnetic fields of the solar wind, stretched along with the wind to the edge of the solar system and spiraled by the rotation of the Sun, rigorously obstruct penetration of galactic cosmic rays in the interior of the heliosphere and to Earth.

The most important property of space weather is its variability. What is essential for us is its quasi-cyclic variability on the time scales of 8÷15 years (the so-called 11-year solar spot cycle). During this cycle the number of spots, expressed in Wolf numbers, and the ensuing activity sometimes falls to zero (minimum solar activity), and then grows to hundreds (maximum solar activity). Simultaneously with the number of sunspots, other manifestations of space weather related to them (solar wind, cosmic rays, magnetic activity of the Earth's magnetosphere) vary as well. On a shorter time scale space weather manifests itself as hours-long magnetic storms, generated by gusts of solar wind, coronal matter ejections, and shock waves in the solar wind. The base of this type of perturbations are the solar flares - an extremely rapid release of magnetic energy in the coronal structures, accompanied by heating of coronal plasma to gigantic temperatures and acceleration of a small group of particles to very high energies (the so-called "solar" cosmic rays). In addition to the variability of space weather on the scales of hours-days (solar flares) and decades (sunspot cycle), there also exist its long-term variations on the scales of hundreds and thousands of years, accompanied by the "stopping" solar activity and a long-term disappearance of sunspots, like the Maunder and Sporer¨-type minima.

---

[4] In fact the prediction of Jevons was partially fulfilled. Dur-ing the 1876–1878 period of minimum solar activity he had specified, the monsoon transfer of moist air from the southwest Indian Ocean towards Southern India ("the jewel in the crown" of the British Empire) has stopped for three years. This led to a rigorous and longstanding drought which caused a humanitarian catastrophe, known as the Great Famine in India: between 6 to 10 million people died of famine, and another 60 million people were forced to leave the disaster region and escape from starvation. The scale of the disaster was to a large extent determined by the decision made by British authorities to abandon massive grain purchases in the unaffected regions of the Empire for the food aid. The decision was taken within the then dominant paradigm of the "government non-interference in the natural economic processes of the free market" and in order to avoid the leap in food prices, predicted by Jevons and capable of initiating the stock market crisis that he had also denounced. Jevons himself got severely depressed, resigned from the London College, and in a short while was found drowned under mysterious circumstances.

[5] This term reflects acute sensitivity of the stock exchange to any "reliable" scientific, economic or political information on the threats of its stability. Such information, which forms the pessimistic expectations of many participants of the exchange trading, can cause real panic, despite the complete unreliability of the initial forecast. The example of Jevons's prediction in particular forms the basis of the concept of the "sunspot equilibria" in the modern economy as a state, taking into account the expectations of players, formed by the a priori information. This information, often being intrinsically irrelevant, may determine the behavior of the players and objectively affect the reality it predicted [8].

[6] This category of space weather also includes the asteroid and comet impact hazards (like the Tunguska and Yucatan), and the nearby supernova explosions, but for our purposes here the effects caused by solar activity are essential.

Let us emphasize that a part of solar activity (e.g., ultraviolet, X-ray and radio emission) varies in-sync with the sunspot number, reflecting the intensity of azimuthal fields in the solar convection zone. At the same time other manifestations of solar activity (coronal holes, flashes) associated both with the poloidal fields, and the nature of convection, often vary out of phase with the number of spots, or with a significant phase shift. As a result, various components of space weather (solar wind, magnetospheric activity, the dynamics of cosmic ray fluxes) demonstrate a complex pattern of phase changes during the cycle [9]. A long-term (tens or even hundreds of years) variability of the relative contribution of different components of solar activity complicates this pattern even more, resulting in possible drastic variations in the phase pattern of different manifestations of space weather.[7]

## 2.2. The Influence of Space Weather on Terrestrial Weather

While the influence of space weather on the Earth's magnetosphere and ionosphere has been acknowledged for about 100 years by now, the possibility of the effects of space weather on the Earth via the changing conditions in the Earth's atmosphere was discovered only recently and is the subject of intense scientific debate to date.[8] The following observations can be listed as a demonstration of such influence.
1) A high correlation between the flux of cosmic rays, penetrating the Earth's atmosphere and the level of cloud cover in some regions, specifically, over the North Atlantic (Figure.1, left) discovered by the Danish scientists Svensmark and Friis-Christensen [10]. Let us emphasize that the observed sensitivity of the Earth's cloud cover to the cosmic rays is in no way universal, taking place always and everywhere in the Earth's atmosphere, but, on the contrary, it is observed only at certain altitudes, in certain geographic areas and over the certain, though lengthy, periods of time [11] (Fig. 1, right panel).

---

[7] As an example, we present the relation of the magnetospheric activity of Earth with the solar activity: if during the first observation period (1868÷1967) the correlation between the geomagnetic activity index "aa" and the number of sunspots "ssn" was highly significant (a 95% confidence interval of the correlation coefficient $R_{95\%} = 0.4 \div 0.72$), then for the next thirty years (1968÷1998) the change of the variation pattern of magnetic activity has led to the disappearance of a significant correlation (a 95% confidence interval $R_{95\%}$ from −0.06 to 0.58). We used the annual data on the number of sunspots from http://sidc.oma.be/DATA/yearssn.dat and yearly numbers (total) of the "aa"-index from ftp://ftp.ngdc.noaa.gov/STP/GEOMAGNETIC_DATA/ /AASTAR/aastar.lst.v12
[8] The discussion of possible contributions of space weather to the so-called "global warming," an abrupt reconstruction of Earth's climate, observed over the recent decades, is particularly piercing.

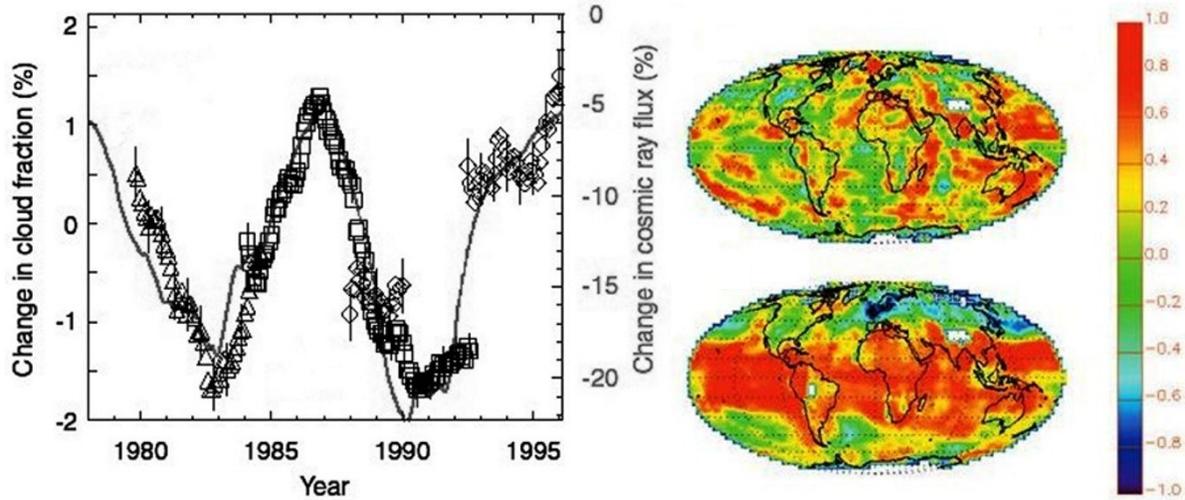

**Fig. 1. Left:** a correlation between the cosmic ray fluxes and cloud cover over the North Atlantic in 1976–1995 discovered by Svensmark and Friis-Christensen [10] (the 1993-1995 data are the subject of debate because of the ambiguity of corrections for the device recalibrations, performed at the time). The left vertical axis describes the variation of the cloud cover area in percentage, the right vertical axis - the variation of cosmic rays flux in percentage, the horizontal axis is time in years. **Right:** a highly inhomogeneous "patchy" distribution of the observed sensitivity of cloud cover (the correlation coefficient) to the cosmic ray flux variation [11]; top - the distribution of correlation with cosmic rays for the local cloud area, bottom - the same for the local temperature of clouds.

Apparently, the main reason for this selectivity is the requirement of fulfillment in the atmosphere in this very place and at this very time of a mandatory critical condition (a "threshold state"), at which the air ionization by cosmic rays can result in significant complementary effects of water vapor condensation and contribute to the observed increase of cloud cover.[9] This pattern is consistent with the latest CLOUD experiment results at the particle accelerator at CERN, where a condensation chamber, simulating the atmosphere is irradiated by high-energy protons. According to the first results of the experiment, the induced formation of aerosols and water vapor on the forming ions and radicals does indeed increase manifold [12].

---

[9] A possible example of such a threshold-type response to the external stimuli is the sensitivity of the water vapor condensation process to air ionization by cosmic rays, which very briskly depends on the density of water vapor: in the regions with an initially low density of water molecules (over the desert zone), no ion additives in the air will lead to the formation of clouds (simply because of lack of water vapor). By contrast, in the regions with the initial excessively high density of water vapor, the condensation and cloud formation occur without any additional contributions of ions in this process. And only when the boundary conditions close to the threshold of condensation are fulfilled, that we can expect a significant effect of the additional air and aerosol ionization on the condensation of water vapor and cloud formation in the given region and at a given time.

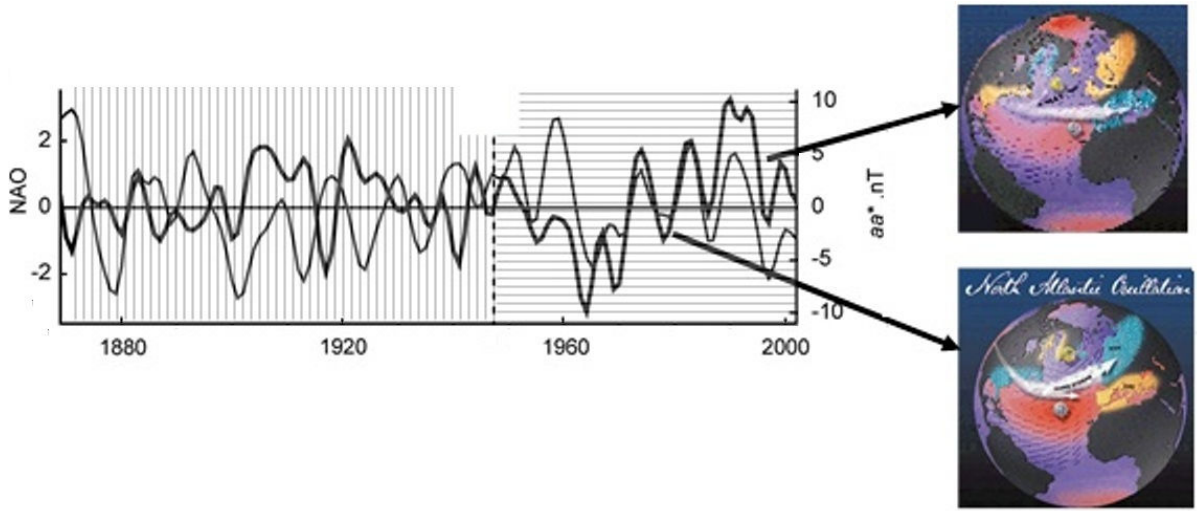

**Fig. 2. Lef**t: the variations of the aa-index (the thin line) and the NAO parameter (the thick line) for the period of 1868–2002. The vertical dashed line (1947) divides the period of the lack of correlation with the period of its appearance. **Right:** the dominant atmospheric circulation at NAO < 0 (top) and at NAO > 0 (bottom) (the illustrations can be found at http://www.ldeo.columbia.edu/res/pi/NAO/).

2) Dependence of the global circulation in the Earth's atmosphere, determined by the NAO parameter (North Atlantic Oscillation), representing the difference between the subequatorial (Azores) and subpolar (Iceland) atmospheric pressures[10] on the level of magnetospheric activity, controlled, in its turn, by the space weather. As shown by the researchers from the St. Petersburg Arctic and Antarctic Research Institute (AARI) [13], a highly significant correlation between the level of magnetosphere activity (the "aa"-index) and the NAO-index variability, which determines the global circulation of the moist Atlantic air (Fig. 2) is observed over the last 60 years, especially in the sub-polar regions (the $P_c$-index).

Note that as in the case of the correlation be-tween the cloud cover over the North Atlantic with the cosmic rays, the detected sensitivity of global atmospheric circulation to the manifestations of space weather showing up as magnetospheric activity is not universal, being implemented only in a certain region and during a certain period of time (since 1946). This selectivity appears to reflect the fulfillment of the same special "threshold" state of the local atmosphere, necessary for the manifestation of sensitivity of the atmospheric processes to the external effects, specifically, formed by the space weather. Convincing evidence of the influence of the solar wind and geomagnetic activity on the NAO-index of global circulation were also presented by Boberg and Lundstedt [14]. Additional evidences regarding the impact of space weather on the climate are given in [15–20]. The possibility of nonlinear effects at the exertion of solar activity on the climate change is considered in

---

[10] At the positive NAO values, the main transfer of moist Atlantic air, carried out by cyclone chains, falls on the central - northern Europe and extends through Russia to Yakutia (with the corresponding contribution to the cloud cover and precipitation in the area), while the Mediterranean region is dominated by the dry and hot weather. Conversely, at the negative NAO values, the cyclone lane moves south to the Mediterranean with the corresponding increase of cloudiness and precipitation there, while a relatively dry weather sets to the north.

[21], and the local property and non-linearity of the response of high-altitude jet streams in the tropopause to the spectral variations of solar radiation caused by solar activity are demonstrated in [22].

An additional contribution to the non-universal nature of the space and earth's weather relationship may be brought in by the variability of the phase portrait of various components of both terrestrial and space weather, sensitive to different manifestations of solar activity and/or the states of the global circulation of the atmosphere. To illustrate this instability we present an evidence of a periodic elimination of the relation between the NAO-index and weather conditions in southern Sweden, although most of the time they are correlated with the changes in the NAO-index [23].

Especially note the approach to the description of the Earth's climate as a system containing not just one state of dynamic equilibrium (the attractor of phase trajectories in the parameter space, describing the system), but several similar attractors [24]. These states are separated by the boundary zones of metastable equilibrium, from where the system can move to an area of another attractor (dynamic equilibrium) under the effects of relatively small external disturbances (solar activity, volcanic emissions into the atmosphere, anthropogenic influence). For such a transition, the phase trajectory has to hit the boundary zone at the time of exposure to the external factors.

## 2.3. The Effect of Terrestrial Weather on the Harvest

The fact that the yield depends on the weather conditions is trivial and well known. However, the threshold nature of this relationship for many cultures is often overlooked. For example, short cold spells during the flowering or heavy rains during the grain harvest could in a few days' time decrease the yields to near zero, and at that not to have a significant effect neither on the yearly average, nor on the monthly average of weather indices (temperature, humidity, precipitation) in the area. Note that for different crops various agronomic and weather conditions may become critical. Thus, for perennial grapes it is important to maintain sufficiently high air temperatures during the maturation, while for the annual crops the soil moisture can play a critical role.

To minimize the crop losses due to the weather anomalies, the practice of long-term selection and zoning of the most appropriate crops, and optimization of agricultural technologies for the weather pattern, dominant in this area are applied. However, exactly this perennial zoning of crops for the "standard" weather in the conditions of a fast and global climate change may lead to the shift of dominant agricultural crops to the state of "risk farming," extremely sensitive to even relatively small and localized weather anomalies.[11]

## 2.4. The Effect of Crop Shortage on the Grain Market

In the free market conditions, the shortages of basic commodity supplies (agricultural products are undoubtedly among them) automatically leads to higher prices, naturally reducing demand towards the equilibrium between the supply and demand. In the situation, where the deficit concerns commodities, included in the list of vital goods (energy, water, food), the market reaction can for a short while take the form of panic with a price hike, much higher than the equilibrium level.[12] This panic reaction of the free market is fully realized in the case of food shortages due to the crop failures resulting from local weather anomalies.

---

[11] Such shifts are also possible in the opposite direction, when climate changes lead to increased productivity and reduce its variability for the traditional cultures of the area. These findings are published in [25], where the impact of the possible climate change on the winter wheat yield in England and Wales is modeled.

It is clear that a sharp increase in food prices stimulates their supply from the external sources that are not affected by the bad weather, which naturally dampens the price leaps and restores the market balance. However, this stabilization mechanism can not always effectively operate: firstly, the final price of the product includes the transportation costs which are sometimes very significant; secondly, in some situations, the freedom of movement of goods is limited by the geographic, economic and political obstacles, like a relative isolation of the market, high customs costs and understated quotas to protect the local manufacturer[13]. Therefore, in the environment of local markets, deprived of the access to external suppliers, or sensitive to the transportation cost growth,[14] the reaction on food shortages, particularly grains, may become panic with an explosive price growth.

## 2.5. Three Necessary Conditions to Implement the "Space Weather–Terrestrial Prices" Relation

Summarizing the above analysis, we can formulate three necessary conditions for the implementation of the cause-and-effect relationship between the space weather anomalies and the grain price hikes they cause.

1) High sensitivity of weather (local in space and time) to the factors of space weather (e.g., cosmic rays, solar UV emission and/or magnetospheric activity).
2) High sensitivity of the leading grain harvests to the weather anomalies (belonging to the area of risk farming) in the given region during the studied historical period.
3) High sensitivity of the grain market to the deficit of supply due to the limitation or absence of the external supply, leading to the explosive (panic) price growth.

## 2.6. The Scheme of the Cause-and-Effect Relationship Between the Space Weather and Terrestrial Grain Markets

Figure 3 presents a diagram showing a possible chain of relations leading to a price response to the adverse conditions of space weather in the regions where the three necessary conditions described above are fulfilled. The key feature of this scheme is the presence of a nonlinear (threshold) sensitivity of the relations between some elements of this scheme (in bold arrows). As a result, the final relation "solar activity-price level" in terms of the catastrophe theory [28] does not look like a "hard" dependence as

$$Y = k_i X_i + \text{Noise},$$

but rather like a "soft"-type dependence

$$Y = k_i(X_i, Y) \times X_i(t - \tau) + \text{Noise},$$

taking into account both the feedback, and the presence of external factors of extrinsic nature, and a possible phase delay. Here X denotes the vector of input variables (the state of solar activity and space weather, the state of the Earth's atmosphere, and the state of the market), Y is the output reaction (price, social characteristics), $k_i$ are the coefficients (functions) of the relations, and $\tau$ is the phase delay.

---

[12] "There is panic buying, and price becomes irrelevant" [26].
[13] Another possible way out of the state of panic price surges for the vital products (food, fuel) are the measures of the authorities on an artificial restriction on the free market via the introduction of the card system for the allocation of products in short supply, depriving the price of the commodity of its equilibrium market functions (the Second World War member countries during and after the war, the Soviet Union for most of its existence).
[14] An example of the effect of transportation costs on the grain market is the convergence of grain price levels in England (the buyer) and the U.S. (the supplier) at the end of the nineteenth century after the mass introduction of freight carriage by cheap steamship lines [27].

## 2.7. Four Possible Scenarios of Price Reaction to Space Weather

Under the proposed scheme for different climatic zones and different types of agricultural production, four alternative responses to the possible local weather modulations by the solar activity/space weather are possible. The modulating factor we use in the proposed set of scenarios are cosmic rays as a possible agent, initiating the formation of additional cloud cover, radiation deficiency, decreased temperature and excess precipitation. Note that a possible relation of the space weather with the terrestrial weather through the cause-and-effect chain "cosmic rays-cloud cover" does not exhaust the list of all possible mechanisms of the influence of space weather on the Earth's weather. In this section we shall limit ourselves to this chain of relations merely to illustrate the possibility of a variety of response scenarios, dependent on the local climatic and agricultural conditions.

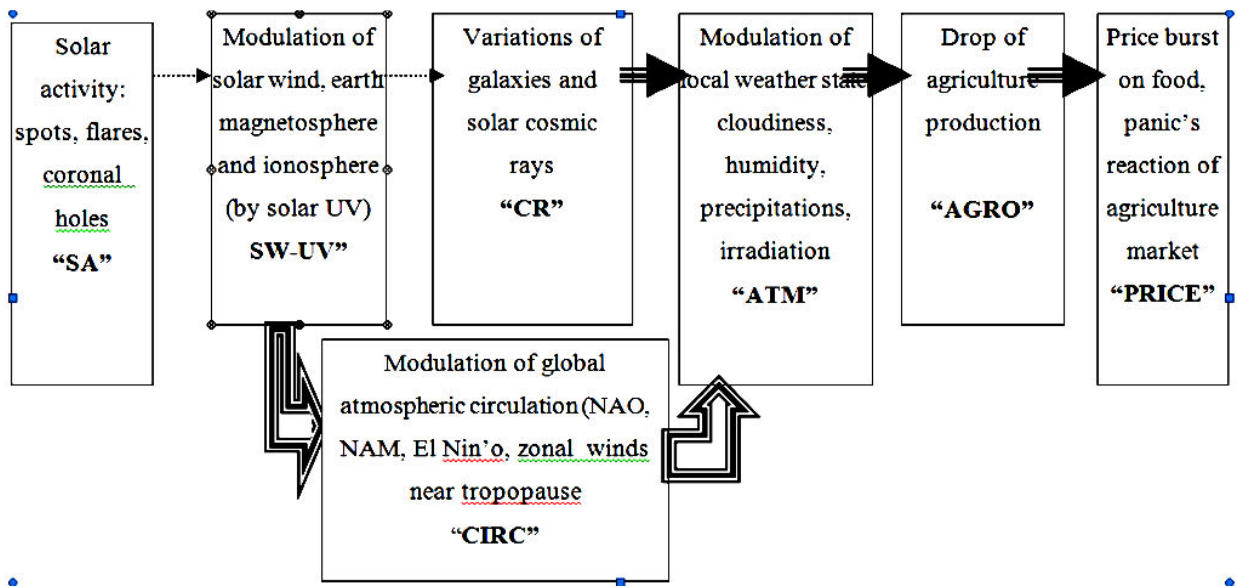

**Fig. 3.** The diagram of the cause-and-effect relationship between the space weather and terrestrial grain markets. The character of the arrows reflects the possibility of the threshold-type reaction: thin arrows correspond to the relations with a direct dependence and without any threshold effects, the bold arrows correspond to non-linear relations with a possible manifestation of the threshold dependence.

All these variants are schematically displayed in Fig. 4 and include the following possibilities.
1) Weather modulations in the zone of risky farming, sensitive to the possible cold spells and excessive rainfall (zone I in Fig. 4). The most likely candidates are the north of Continental Europe, England. The most unfavorable phase of solar activity is its mini-mum with the minimum intensity of the solar wind and the maximum flux of cosmic rays that enhance the formation of clouds in the sensitive areas (e.g., the North Atlantic (Fig. 1, right)).
2) Weather modulations in the zone of risky farming, sensitive to hot weather, hot winds and droughts (zone III in Fig. 4). The likely candidates are southern Europe and the Mediterranean (Italy, Spain, North Africa). The worst phase of solar/space weather is the solar activity maximum and solar wind with a deficiency of cosmic rays and a decrease in their contribution to the formation of clouds over the Atlantic.
3) In some cases, there may appear some areas of risk farming, both sensitive to colds and excessive precipitation, and to a shortage of rainfall and drought. This situation can be the case for the areas with climates, particularly unfavorable for the agriculture, allowing successful agricultural production in a very narrow range of climatic parameters (precipitation, temperatures, humidity and solar radiation). This situation is described by zone IV in Fig. 4. The likely candidate is Iceland in particularly unfavorable times of climate change.
4) There may be situations where the agricultural production is not at risk farming and is by definition weakly sensitive to the changes in weather conditions in the region. This situation is described by zone II in Fig. 4 with a neutral reaction of agricultural production to the space weather/solar activity phases.

These four types of reaction exhaust all possible scenarios of the relations between the space weather and terrestrial agricultural production for the expected type of influence "cosmic rays–cloud cover." The further analysis will have as its aim the search of the implementations of these specific scenarios in certain geographic regions and during specific historical periods.

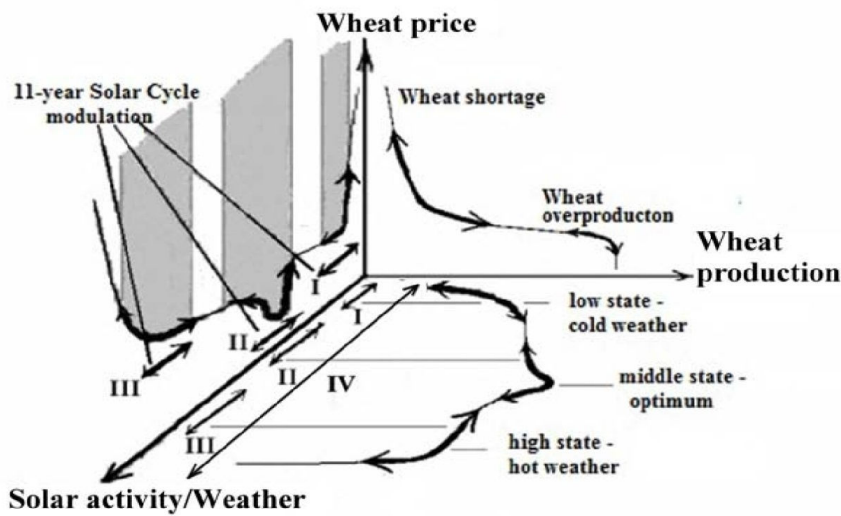

**Fig. 4.** Possible reactions of the agricultural production and market to the modulation of local weather by solar activity/space weather.

According to the above description, we can expect systematic price leaps in the states of the solar activity minima for the cold and humid regions (zone I in Fig. 4). Similar price hikes for dry and hot regions (zone III) are expected for the states of solar activity maxima. For these two cases, the typical periods between the price leaps are determined by the period of solar activity (9–13 years). For the regions sensitive both to the excessive rainfall, and precipitation shortages (zone IV) price hikes are possible both in the solar minima and maxima with the characteristic periods of a half of the solar cycle, (4÷6 years). The same period variations can result from the artificial mixing of market data from the areas of opposite types of price reaction into a single sample, like, for example, in Beveridge [29], who used the "European mean" price index as a parameter, and obtained a significant Fourier-transform response exactly at the periods of 4–5 years.

## 3. DATA AND SELECTION OF RELEVANT ANALYSIS METHODS

### 3.1. The Data Used

#### 3.1.1. The Solar Activity Data

The main kind of data on the solar activity (the generator of space weather) is the numbers of sunspots. Although their observations began in 1611 straight after the Galileo's discovery, however, the Maunder minimum that followed has stopped the observations of spots on a regular basis. Regular daily observations of sunspots by several groups of researchers had begun only after the restitution of solar activity in the early 18th century. It made it possible to create a catalog of solar activity (the number of sunspots) from 1700 to the present time.[15] However, the data for the previous period have also been recovered from the years when at least some manifestations of spots on the surface of the Sun were recorded.[16]

Another method we used to estimate the solar activity during the Maunder minimum and preceding years was an indirect method of estimating the level of solar activity from the abundance of isotopes $Be^{10}$ in the Greenland ice cores [31]. Though this method does not give a reliable quantitative estimation of the actual number of spots, it allows to relatively reliably determining the times of solar activity maxima and minima. Note that for our purposes, the data on the $Be^{10}$ isotopes, directly reflecting the intensity of cosmic rays penetrating the atmosphere, is much more important than the number of sunspots. Cosmic rays have a direct influence on the atmosphere, its ionization and condensation of water vapor, while the causal chain from sunspots to atmospheric processes includes many intermediate elements (the solar wind, the state of the Earth's magnetosphere and global atmospheric circulation). These factors can mask a possible link between the atmospheric manifestations and actual solar spots.

---

[15] http://sidc.oma.be/DATA/yearssn.dat
[16] The results are discussed in [21, 30] and presented at the NOAA/NGDC website: ftp://ftp.ngdc.noaa.gov/ /STP/SOLAR_DATA/SUNSPOT_NUMBERS/ANCIENT_DATA/ /earlyssn.dat

### 3.1.2. The Data Sources on the Prices of Grain (Wheat)

The first data on the prices of grains used by William Herschel [3] for his analysis were published in the above-mentioned work of Adam Smith, called An Inquiry into the Nature and Causes of the Wealth of Nations [2]. However, the most complete and accurate database of grain prices in the medieval England was first compiled as a result of the selfless work of an outstanding economist and statistician Professor Rogers [32]. His database covers the period from 1259 to the mid-18th century; uses the data on the prices of purchased grain in the monasteries and colleges, which at that time were the centers of literacy and accounting; this fact increases the reliability of sources of prof. Rogers.

In addition to the data on wheat prices, we used the data on the price of the consumer basket over the period of 1264–1954 [33] for the analysis.

An additional source of data that we used was the archive of wheat prices for 90 cities of Medieval Europe. In particular, we used the most complete part of data for the time range of 1590–1702, covering the period of the Maunder minimum and the interval of observations of the Be10 isotope variations [31]. These price data were adopted from the database of the International Institute of Social History [34].

In addition, to search for possible manifestations of the influence of space weather on the wheat markets in the early modern period, we used the U.S. Department of Agriculture data for the period of 1866-2002 on the prices for durum wheat, used for the production of bread and bakery products.

We have also used a comparative analysis of the production and prices for grain in the medieval England and France [35], the analysis of grain prices in the medieval England from the aforementioned study of Beveridge [21], the analysis of the correlation between the weather and crop yields in the late 19$^{th}$ - early 20th centuries [36], the analysis of grain production in the U.S. of the early 20th century [37, 38].

To analyze the effect of the feedstuff failures in the 18$^{th}$ - 19th century Iceland on the livestock reduction and the resulting hunger and depopulation, we used the results of Vasey [39].

## 3.2. Methods of Analysis

The main difficulties in finding the responses of agricultural product prices, used as an indicator of a possible relation with anomalous space weather conditions are caused by the two following circumstances.

1)      The solar activity cycle, as the main generator of space weather variations is not stable neither in the frequency of sunspot number variations, nor in the amplitude of the cycle. The temporal interval between the minima of solar activity (the cycle period) varies in a wide range of periods from 8 to 15 years. The amplitude of the cycle, expressed in the so-called Wolf numbers, ranges from hundreds to tens of years (sometimes even dropping to zero for dozens of years). Solar activity includes various components (spots, coronal holes, flares), formed by different dynamo process elements or the combinations thereof (poloidal and azimuthal fields inside the Sun, convection, differential rotation and meridional circulation). As a result, phase changes of various components of solar activity are very different during the cycle period. For example, the contribution of coronal holes and resulting recurrent fluxes in the solar wind is maximal during the sunspot minimum, though nonexistent during the sunspot maximum. Solar flares of a very small amplitude correlate well with the number of spots, whereas for the powerful solar flares with the emissions of coronal plasma and accelerated high-energy protons, the situation is much more complex and dynamic - from cycle to cycle the distribution of proton flares dramatically changes. In a number of cycles their frequency is maximal not at the sunspot activity maximum, but in the phase of growth or decline of the number of spots [40]. Instability of the relative contribution of different components of solar activity in the formation of space weather creates additional complexity, e.g., a variation of the relative contribution of coronal holes and sunspots in the disturbances of the Earth's magnetosphere, caused by the specificity of the solar dynamo process [9, 41]. The influence of the factors of space weather on the atmosphere is also taking place against the background of a complex, and not fully understood today pattern of global atmospheric circulation with long-range effects,[17] phase instability and possible strange attractor-type transitions.

2) The influence of space weather on the behavior of prices on agricultural products (grains), if it would indeed exist, has to take place against the background of other effects, simultaneously in force. These effects can have comparable amplitude and both a random nature, or, possibly, periodic components at the same temporal ranges. Among such factors we may list the climatic variations, political and military events leading to the economic shocks, as well as the scientific and technological revolutions.

This situation renders inefficient the use of classical statistical methods, both aimed at the identification of the harmonic signal (Fourier analysis, periodogram analysis) with the given (for example, an 11-year) period, and the search of a direct linear relationship (correlation and regression analysis). It requires the use of other methods and statistics, more robust with respect to the variability of the period and amplitude of solar activity, instability of the atmospheric circulation and other external factors.

Under the proposed approach, such statistics and methods of analysis may be the following:
1) a comparison of the statistics of intervals between the grain price leaps with a similar statistics of intervals between the extreme states of solar activity (e.g., its minima);
2) a possible systematic asymmetry in grain prices at the "favorable" and "unfavorable" states of space weather/solar activity;
3) regression analysis using dummy 0/1-type variables, identifying the system status in the binary form.

---

[17] As an example, we present a situation where extreme weather conditions related to the El Nino cycle of the west coast of the South and Central America through a series of atmospheric and oceanic transfers lead to a dramatic weather response at the opposite end of the globe in the North Atlantic Ocean. Note that the researchers indicate a possible link between the solar activity and the El Nino/LaNina effects shaping the climate anomalies [42].

For the analysis of specific situations in the given regions, we have to consider the possibility of the above-mentioned "long-range" effects. This term describes the fact that the regions that are remote from the effective influence of space weather on the terrestrial weather, can, however, prove to be sensitive to these modulations as a result of the global atmospheric circulation and cyclonic transfer of the formed cloud cover over thousands of kilometers (e.g., from the North Atlantic to the Eastern Siberia).

Another important effect which should also be considered in the analysis is the possible phase delays of the price response to the adverse effects of space weather. The reasons for such delays may be the presence of grain reserves from previous harvests, a mismatch of calendar and agricultural years in the statistical reporting, inertia of agricultural markets.

The sensitivity to weather conditions has to manifest itself more fiercely in the situations with grain production, concentrated in a compact region (up to hundreds of kilometers) with one and the same type of weather conditions, rather than for the agricultural production, distributed over thousands of kilometers with different climate conditions in various regions and, consequently, with different (even to the opposite) types of sensitivity to the external factors.

We took into account these very parameters to study the data over several regions, registered during different historical periods.

## 4. RESULTS OF ANALYSIS
### 4.1. Sensitivity of the Wheat Market in Medieval England to Space Weather

Medieval England is a perfect testing ground for the search of the effects of space weather on the grain prices, since this very region and at this very historical period satisfied all the three above conditions necessary to bring into action such a relationship:

1) the dependence of weather conditions on the factors of space weather in the region of cloud formation over the North Atlantic, sensitive to the cosmic-ray variations during the changes in solar activity;

2) belonging to the zone of risky farming (particularly for wheat), extremely sensitive to the adverse weather anomalies during the growing season;

3) relative isolation from the European markets enhancing the price reaction of the local market for the grain shortages.

Another advantage of the study of this very region is the presence of the above mentioned highly reliable data archive on the grain prices from 1259 up to the eighteenth century, compiled by prof. Rogers [32]. The initial curve of variations of the yearly average grain price is presented in the top plot of Fig. 5.

To search for the manifestations of the space weather effects, we used the above described techniques for comparing the interval statistics and looked for the developments of price asymmetry.

First, we compared the statistics of intervals be-tween the bursts of wheat prices with the statistics of intervals between the solar minima - both for the distributions of these intervals by duration (Fig. 5, bottom plot), and for the statistical characteristics of these distributions (comparing the average intervals of these two samples, their medians and dispersions) [43, 44].

As we can see from the table of statistical parameters of the interval distribution, just like from their histograms (Fig. 5, bottom left), the distribution of intervals for the grain price bursts and the consumer basket coincide with the distribution of intervals be-tween the solar activity minima with a high level of confidence (99%).

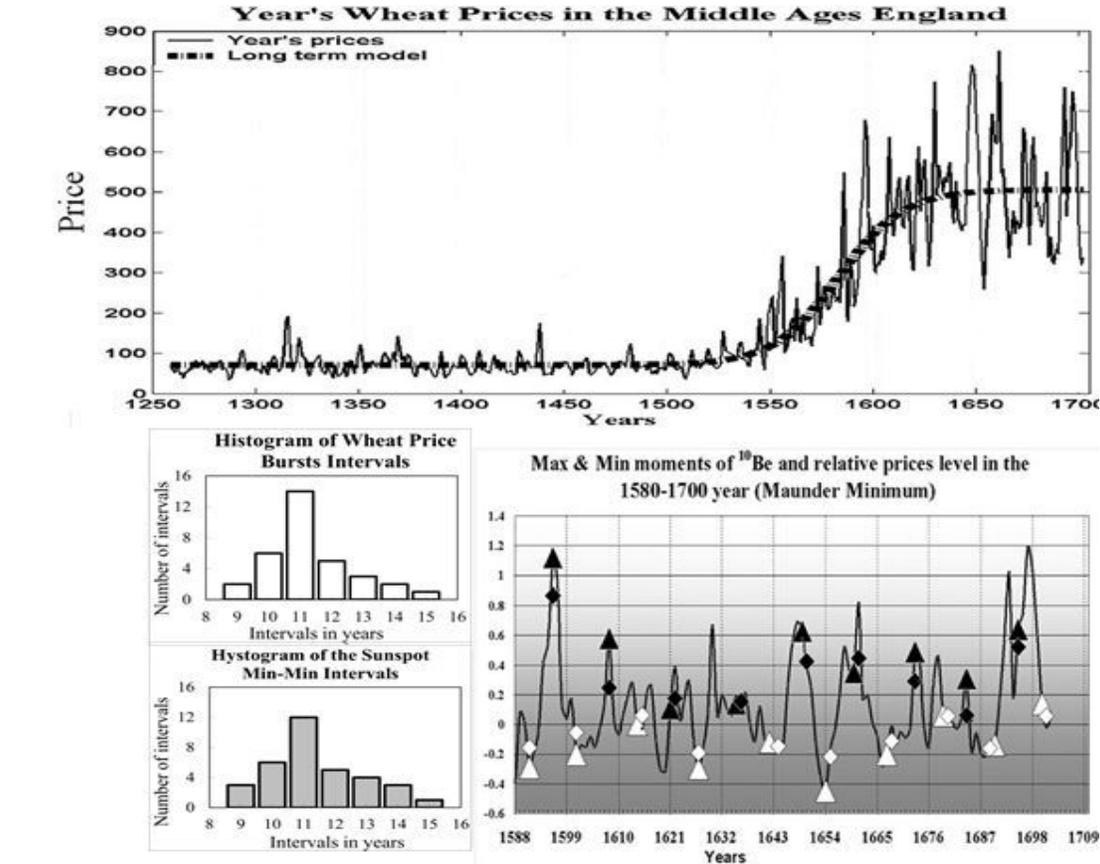

**Fig. 5. Top plot:** wheat prices in the Medieval England, smoothed by a logistic function. **Bottom left plot**: a comparison of histograms of the intervals between the solar activity minima and the intervals between the wheat price shocks. The range of intervals along the horizontal axis: 8-16 years with an increment of one year; the maximum for both histograms falls on the interval of 11 years. **Bottom right plot**: price asymmetry at the points of solar activity minima/maxima, determined by the level of $Be^{10}$ in the period of the Maunder minimum: the vertical axis describes the difference between the yearly prices and the yearly prices averaged by the logistic curve, normalized to the average values; white triangles mark the prices during the peak year; the white diamonds mark the average prices for three years around the peak year; the black triangles and diamonds mark the same for the periods of solar activity minima.

Another proof of the influence of space weather on wheat prices in the medieval England is a highly significant price asymmetry during the minimum and maximum solar activity states in the Maunder mini-mum period, given in the bottom right plot of Fig. 5. As we can see from the graph, all the nine cycles of solar activity during this period are characterized by systematically excessive wheat prices in the years of solar activity minimum, as compared with the prices during the next maximum[18]

---

[18] To identify the states of solar activity minima and maxima during the period of Maunder minimum, we used the data of prof. Beer's team [31], who had reconstructed the level of cosmic rays and solar activity during the studied period based on the abundance of the isotope Be10 in the ice sheets of Greenland (see the details in [43]).

(on the average by two times). The level of significance in this case is in excess of 99%.

The above results demonstrate the reality of the implementation of the cause-and-effect chain between the space weather and terrestrial grain markets for the specific case of medieval England. The observed type of market reaction corresponds to the variant I (zone of risky farming, sensitive to the possible cold snaps and excessive described in Section 2 of this article, what is in good agreement with the expected type of reaction for this climate zone.

**Table 1.** A comparison of statistical parameters (median, mean, dispersion) for two samples of intervals between the price shocks (for the wheat prices and for the consumer basket cost) and for the sample of intervals between the minimal states of solar activity. An agreement between the results obtained from the wheat prices, and consumer basket costs reflects the fact, trivial for the given historical period, that the basis of the consumer basket cost at the time was the meal expenses, where the bread and cereal products made up a very substantial share

| Samples used | Median | Average | Standard deviation |
|---|---|---|---|
| | | | |
| *Min-Min intervals for sunspots* | *10.7* | *11.02* | *1.53* |
| *Intervals between grain price shocks* | *11.00* | *11.14* | *1.44* |
| *Intervals between the price shocks of the consumer basket* | *11.00* | *10.50* | *1.28* |

## 4.2. Sensitivity of Wheat Markets in Medieval Europe to Space Weather

In the following stage of the analysis, we tried to answer the question whether the observed sensitivity of the grain markets in Medieval England is some kind of a universal feature common around the globe (similar to the dependence of these prices on the winter-summer season change), or this property of the markets occurs only in the isolated regions and during the given historical periods, when these areas more or less fulfill all the three conditions necessary for the implementation of the causal "space weather–terrestrial prices" relation (Section 2.5 of the present paper).

To answer this question, we have conducted a regression analysis of the data from 22 European grain markets, possessing relatively complete data among all the markets, presented in the database of the International Institute of Social History [34].

For the analysis, we used the regression analysis method with "dummy" variables, introduced in practice by D. B. Suits [45][19]. In our study we have used this method to establish the connection between one or another state of the grain market with the states of minimum or maximum solar activity. We tested the hypothesis of the presence of this relationship between the prices and solar activity, and estimated the statistical significance of this relationship for the samples of markets from different European regions. For this purpose, we have introduced a "dummy" variable $d_{min}$, to which the value 1 is assigned for the years of solar activity minimum and value 0 - for all other years. Similarly, we used a "dummy" variable $d_{max}$ for the years of maximum solar activity. In a sense, this method is the development of the technique of direct search for the price asymmetry in the minimum and maximum solar activity phases, used above for the wheat market in medieval England (Fig. 5, bottom right plot). However, the method of "dummy" variables is more accurate and allows obtaining a quantitative estimation of the significance level of the claim that such a phase asymmetry is indeed present.

---

[19] "Fictitiousness" of these variables is only in that they quantitatively describe the qualitative characteristics of the "yes"–"no" type.

For our analysis we have used the historical period of 1590–1702 with the Maunder minimum (a period of a sharp drop in solar activity), imposed on it. The choice of this period for our analysis was determined by two circumstances:
1) during the chosen period the climate of a considerable part of Europe was experiencing the "Little Ice Age" with the transition of significant areas of agriculture to the state of "risk farming" and, accordingly, an increase of the role of weather anomalies in the production of grains;
2) for this period (and not only for it) there exist especially conducted measurements of the Be10 isotope abundance, most directly reflecting the contribution of cosmic rays (one of the possible factors influencing the weather conditions) [31][20]

Figure 6 presents the map of Europe, which shows the amount of annual precipitation. We report the significance levels of wheat price hikes during the minimum and maximum solar activity conditions for a number of markets, where this response was highly significant. We can see from the figure that all the grain markets of England during this period have shown a high sensitivity to the states of space weather, related both to the minima ($d_{min}$), and maxima ($d_{max}$) of solar activity (with a very high confidence level, reaching 99% and above). Besides them, a high response significance level (over 95%) to the solar activity minima was shown by the wheat market in Leiden (the Netherlands), located in similar climatic conditions. At the same time some markets in Belgium, France, and especially in Italy (Naples, Bassano del Grappa) demonstrated a significant sensitivity to the solar activity maximum. Since one of the channels of the potential influence of space weather on the terrestrial weather is the modulation of cloud cover in the northern Atlantic by a flux of cosmic rays, we can expect that during the solar minima, excess rainfall and a lack of solar radiation may appear (unfavorable for the agriculture in the cold and humid areas). At the same time, in the state of maximum solar activity, the flux of cosmic rays and the associated formation of clouds are declining;

---

[20] An additional analysis, performed using the regression analysis with "dummy variables" for different temporal intervals, and not used in this article, shows the presence of a significant relation between the solar activity extrema and wheat prices in England of the period until the 1840s. We explain the weakening of the market sensitivity to the factor of space weather after this period by a sudden increase in the contribution of imported grains - from 5÷8% in 1801–1840 to 17÷78% starting from 1841 [27]. This increase in the imports broke the causal chain in the link "isolated market sensitivity to the supply shortages."

hence, there may appear the drought conditions, unfavorable for the agriculture in the southern regions of Europe, being under the effect of a hot and dry climate of North Africa, especially in Italy and Spain.

The listed results show that the observed sensitivity of wheat markets to the factor of space weather/solar activity is not a universal property, constant for all regions and historical periods. On the contrary, depending on whether all the necessary conditions stated above (Section 2.5 of the paper) are fully or partially fulfilled, either there appears no reaction to the factor of space weather/solar activity (for most of markets in Central Europe), or a highly significant sensitivity to the solar activity minima (England and neighboring European markets), or a notable response to the solar activity maxima in the regions sensitive to drought (e.g., Italy, located partially under the influence of dry and hot climate of North Africa). Thus, the observed distribution of the zones of sensitivity to space weather, as well as the sign of this sensitivity are in good agreement with the above scheme of the possible cause-and-effect relationships between the space weather and terrestrial grain markets.

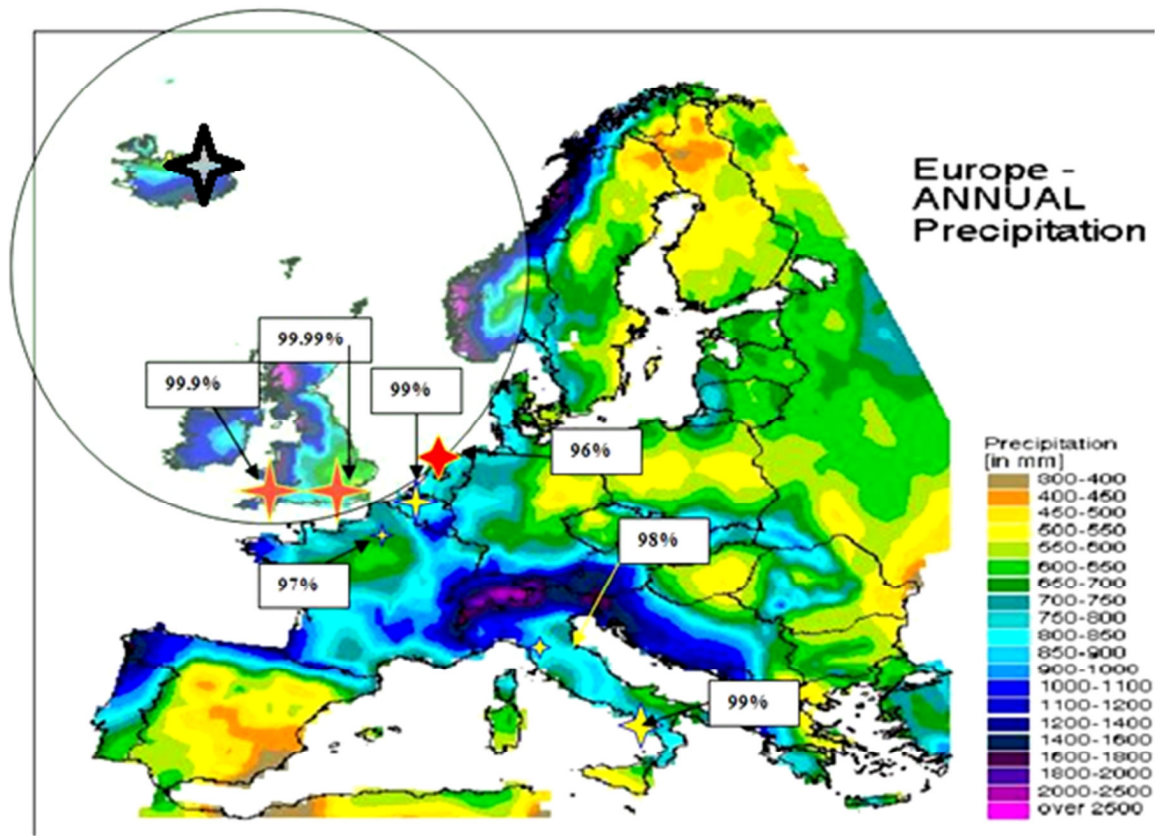

**Fig. 6.** The Europe map of precipitation presented climate zones distribution. Localization of wheat markets that show high sensitivity to the space weather and its significance level, are shown. Red stars show markets sensitive to states of minimal solar activity (London, Exeter – England; Leiden – the Netherlands), yellow stars show markets sensitive to states of maximal solar activity (Napoli, Bassano – Italy). The stars size corresponds to the significance level.

## 4.3. Space Weather Sensitivity of the U.S. Wheat Market in the 20th Century

The entire previous discussion referred to the grain markets in the Middle Age or the very beginning of the early modern period. The question naturally arises whether the above-described scenario of the effect of space weather on the terrestrial markets is realized anywhere on the planet in the contemporary history as well.

At first glance, the widespread introduction of modern farming methods, increasing the resistance of plants to adverse weather conditions should violate the second condition (belonging to the area of risky farming, Section 2.5), necessary to implement this scenario. Another factor of suppression of the possible sensitivity of grain markets to external anomalies is the globalization of the world economy in the 20th century, dramatically facilitating the market access for foreign suppliers. However, we decided to analyze the U.S. grain market in the modern period for the search of possible manifestations of the dependence of grain prices on the space weather. For this analysis, we used the wheat prices over the period of 1908–1993 according to the USDA data [46].

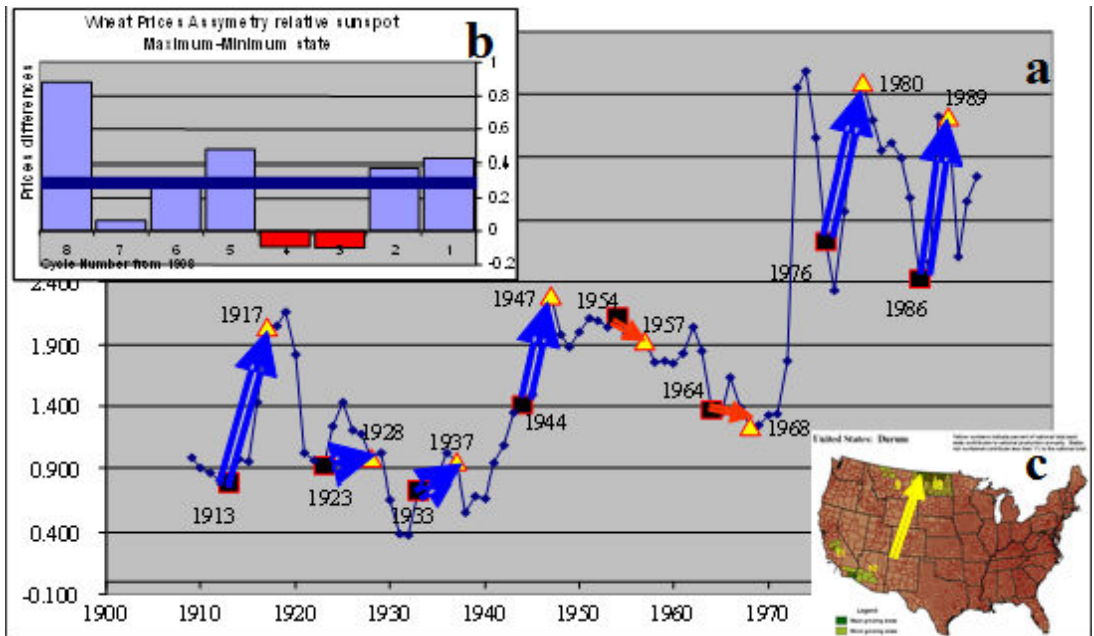

**Fig. 7**. **a** – Changes in durum wheat prices in USA in the period 1908-1993. Blue points and lines show price dynamics, yellow triangular - years of solar activity maximums, black squares – years of solar activity minimums, blue double arrows – price changes from a solar activity minimum to the next maximum. **b** – histogram of price asymmetry = a price difference between the maximum and the predecessor minimum of solar activity – depending on number the solar activity cycle. **c** – high concentration of the durum wheat crop: more than 2/3 of the harvest area is concentrated in 3% of the USA area, in the state of North Dakota.

Figure 7 demonstrates the wheat price fluctuations for the above period. A small sample size, including only 8 cycles of solar activity does not allow us to investigate the statistical properties of the intervals between the price hikes, as it was done for the medieval England based on the 500-year-long statistics. In this situation we can only hope for the manifestation of the phase asymmetry in prices, creating a systematic and significant difference be-tween the prices of grains in the states of solar activity maxima and minima. As we can see in Fig. 7, for the sample used there indeed exists a significant systematic price asymmetry during the periods of solar activity minima and maxima: the prices in the states of solar maxima are systematically higher than the prices in the previous solar activity minimum. To estimate the level of statistical significance of this asymmetry, we used the Student criterion. At the average asymmetry of ΔPrice = 0.29, and the standard error of the mean s(ΔPrice) = 0.12, the level of significance for rejecting the null hypothesis on the absence of asymmetry exceeds 97%. Therefore, for the studied sample of wheat prices in the USA we can argue that even during the Contemporary history there still remains a significant influence of space weather on the agricultural markets in the USA, although the significance level of such effect is somewhat lower than that observed during the Maunder minimum[21] (Fig. 5c). This unexpected result may be due to the fact that the durum wheat seeding in the USA is concentrated in a very compact area: about 70% of all the seeding are located on the 3% area of the U.S. state of North Dakota at the border with Canada - in the area severely affected by North Atlantic Oscillations (NAO), in turn, sensitive to the effects of space weather.

### 4.4. Possible Manifestations of Space Weather Anomalies through the Local Anomalies of Terrestrial Weather in the Famine Statistics

One of the most tragic manifestations of abrupt biting price hikes, caused by the crop failures due to adverse weather conditions is the famine and the concomitant decline in the population due to both a higher mortality and mass migration/emigration from the affected areas. The above influence of space weather on the terrestrial climate, which manifests itself in the form of crop failures and grain price fluctuations, should also, from general considerations, leave its mark in this sad statistics. Obviously, such a manifestation of space weather can only take place in certain regions, and only at certain periods for which all the three necessary conditions of the "space weather - agricultural product prices" are fulfilled (Section 2.5).

We have chosen the nineteenth century Iceland as a possible region for the search of such manifestations. The following reasons made us select this very region and this very period of time:
1) The location of Iceland in the region of North Atlantic, highly sensitive to the factors of space weather ("cosmic rays–cloud cover," the North Atlantic Oscillation), which meets the first requirement from Section 2.5.
2) Belonging of the 19th century Iceland to the area of risky farming due to the specificity of the agriculture of the day, extremely sensitive to the weather conditions. During the period under study, the main source of food was not the coastal fishing, as at a later time, but livestock, totally dependent on the grass harvest on local pastures. The herbs of Iceland are considered among the best in Europe owing to the long solar day in the short period of cool summer. Thus, the rural economy of the country was extremely sensitive to the weather conditions (freezing weather, clouds, the share of sunny days, precipitation), and through them - to the space weather. Note that in the situation with Iceland it is both the excess of rainfall and lack of solar radiation, as well as the rainfall deficiency and drought (a situation similar to the case IV in Fig. 4) that could be responsible for the feedstuff crop failure. Therefore, for Iceland of the time we should expect a negative response to both states of the extreme solar activity, its minimum and maximum.

---

[21] About the same significance level of the "solar activity–prices" relationship (96%) allows to apply for this sample the "dummy variable" method, described in Section 4.2.

3) The isolation of Iceland during the studied period from the basic food markets, the supply from which could mitigate the effects of crop failures (the third condition in Section 2.5).

The work of an American researcher D. A. Vasey [39] is the source of data on the crop failures and the resulting cases of livestock reduction and human depopulation. As shown in this work, all the periods of population decline in the 19th century Iceland coincide with the reduction in the number of livestock, caused, in its turn, by the reduced feedstuff harvest owing to the abrupt weather anomalies (extremely severe winters with anomalously long periods of winter storms and low temperatures). The corresponding periods of famine caused by the livestock reduction due to the lack of feedstuff are marked with black squares in Fig. 8. It is striking that the onsets of all the six periods of depopulation (large-scale markers) always either coincide or fall on the years closest to the beginning of an extreme state of solar activity (sunspot minimum or maximum), lasting from one to three ÷ four years.

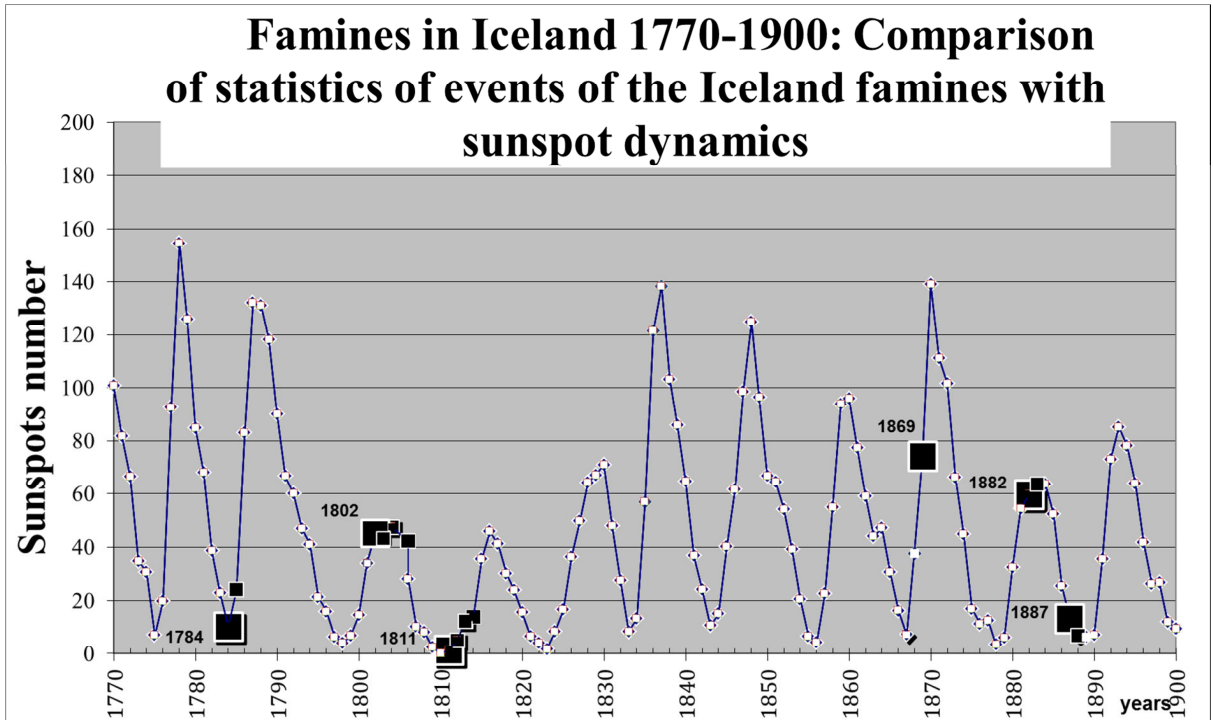

**Fig. 8.** The relationship between the solar activity and periods of depopulation in the late 18th–19th century Ice-land due to the livestock reduction, caused by the weather anomalies against the phase curve of sunspot variations. The black squares mark the periods of decreased popu-lation, where the large squares mark the times of famine onsets.

To test this hypothesis, we have identified the periods around these events - the "locust years" secreted from one minimum to the next one for the events that occurred in the period of the cycle maximum, and, conversely, from one maximum to the next one for the events that took place during the cycle minimum. To reduce the data from different cycles, varying by duration, amplitude, and shape (asymmetry) in a single homogeneous sample for each year, the number of sunspots was renormalized in relative amplitude, while the points of time - to the relative cycle phases. For the events around the maximum, the normalized amplitude $y_{norm}$ was determined as a deviation from the minimum value of the number of sunspots in the given cycle, normalized to scale of cycle variations:

$$y_{norm} = (ssn_i - ssn_{min})/(ssn_{max} - ssn_{min})$$

where $ssn_i$ is the number of sunspots per year $i$, $ssn_{min}$ is the least number of sunspots per year of minimum, $ssn_{max}$ is the number of sunspots in the year of maximum.

For the events in the vicinity of the cycle minimum we have used the absolute deviation from the maxi-mum value normalized to the scale of the spot number variations in the cycle:

$$y_{norm} = (ssn_{max} - ssn_i)/(ssn_{max} - ssn_{min}),$$

The points in time $t_i$ were recalculated to the phases of cycle $\Phi_i$ with the following formula:

$$\Phi_i = |t_i - t_{ext}|/(T_{cycle})$$

where $t_{ext}$ is the time of extremum (minimum or maximum) of the given cycle, and $T_{cycle}$ is the duration of corresponding cycle in years. The resulting normalized phase data for the six cycles during which the decreases in population were noticed due to the feedstuff failure and reduced live-stock [39] are united in a sample shown in Fig. 9.

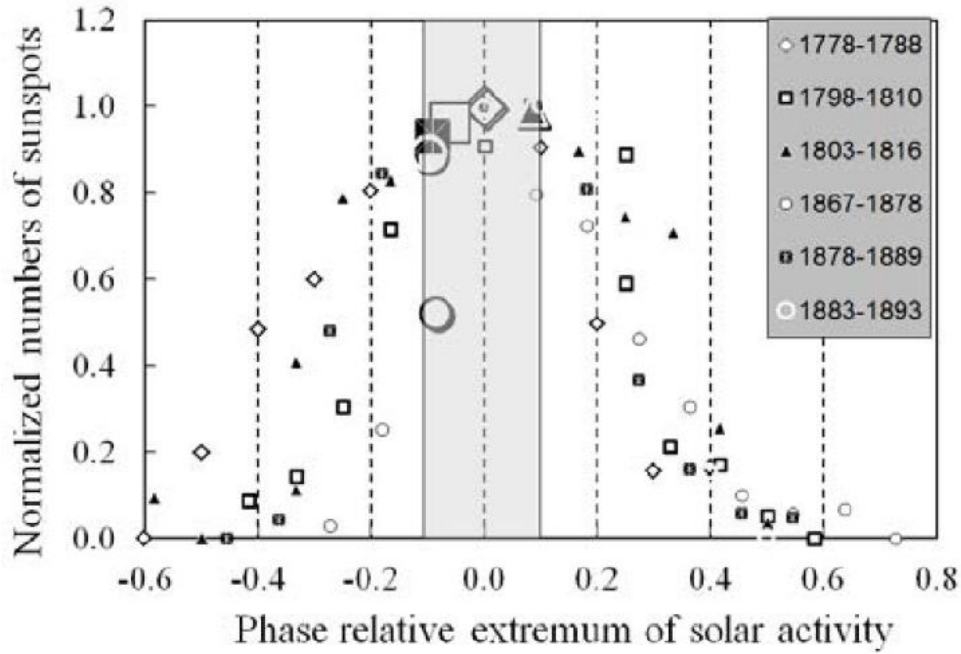

**Fig. 9.** The phase diagram, formed by imposing six periods of solar activity, during which there has been a decline in the population and reduction of livestock number in Iceland. The horizontal axis shows the phase of the so-lar activity cycle relative to its extremum (maximum or minimum), small signs display the normalized number of sunspots for different cycles (time intervals for each sequence are given in the sign description), large signs mark the years of onset of the population decline for the given cycle. The gray rectangle in the center of the figure limits the fraction of the phase cycle space, in which the depopulation events took place.

As we can see from this figure, all the six cases of depopulation in Iceland during the period of 1784÷1900 fall on $\Delta\Phi_0 \leq 20\%$ of the phase space of the solar cycle around the cycle extremum (maximum or minimum). Given that our reference of famine events to the solar activity extrema (i.e. both to the minima and maxima) narrows the scope of the independent phase space by half, the recalculated fraction of the phase interval, which contains all the investigated cases of hunger, has to be reassessed as $\Delta\Phi_f = 2\Phi_0 \leq 0.4$ from the duration of the whole cycle. The corresponding probability P of a statistically distributed realization of a match for six independent events, randomly and uniformly distributed over the phase space of the cycle, can be estimated as $P = 0.4^6 \approx 0.004 < 0.5\%$, which, in our view, excludes the null hypothesis of the random nature of this coincidence.

The observed phase locking of the depopulation periods in Iceland owing to the crop failures, caused by weather anomalies, to the solar activity extrema allows to assert the high reliability of the manifestations of a negative effect of space weather/solar activity extrema on the state of agricultural market in this region during the studied historical period. This type of reaction is consistent with the response, expected for this region according to the scenario IV based on the classification in Fig. 4.

## 5. CONCLUSIONS AND DISCUSSION

This paper describes the model of a possible influence of space weather on the terrestrial agricultural markets, based on the cause-and-effect chain "space weather–terrestrial weather–agricultural production–market prices."

1) We noted that the implementation of this chain requires a simultaneous execution of a set of necessary conditions in the given place and at a given historical period. Therefore, the relationship between the space weather and the state of agricultural markets is not universal. It manifests itself only in those areas and only during those historic periods, where and when all the necessary conditions (atmospheric, agro-climatic and market) are fulfilled.

2) We have demonstrated a reliable manifestation of this dependence for the case of medieval England (especially during the Maunder minimum).

3) We have shown that the sensitivity of the European grain markets to space weather (including the response sign) depends on the localization in the corresponding climate zone.

4) We have demonstrated that even in the Contemporary history the effect of the dependence on space weather/solar activity persisted in the U.S. for the durum wheat, produced in a very compact area in the zone, sensitive to the influence of the effects of the North Atlantic Oscillation.

5) We have illustrated the coincidence of the periods of depopulation in the 18th–19th century Iceland, induced by famine due to the feedstuff failure and livestock reduction owing to the weather anomalies with the extreme phases of solar activity/space weather. We explain this coincidence as a sign of the dependence we study between the space weather and harvest/prices.

The results we presented demonstrate the substantiality of the effects of space weather manifesting themselves in the behavior of terrestrial markets of agricultural products for the regions which fulfilled the necessary conditions described above to trigger the corresponding causal chain in the studied period.

A constant development of agricultural technologies employing the achievements of genetic engineering, biotechnology, the use of artificial irrigation and an active introduction of agricultural chemistry and methods of plant protection has to lead to an increased stability of crops to the external factors such as weather anomalies. This process should automatically break the cause-and-effect chain "space weather-agricultural product prices," suppressing a key section of this chain, "weather–harvest." Therefore, in the near future we may suggest a gradual weakening of the sensitivity of prices of grains and other agricultural products to the factor of space weather.

Unfortunately, the global and drastic climate change observed over the last years could lead to quite opposite results. The main feature of this change is the dramatic increase in the number and amplitude of weather anomalies, particularly the drift of traditional temporal boundaries of seasonal weather conditions. In these conditions, in a large number of regions a decades-long (and even centuries-long) selection of crop varieties most appropriate for the "standard weather" of the region may become non-optimal. A severe departure of local weather conditions from the local "standard" may result in a transition of a significant number of regions to the state of risk farming, when agricultural production would be highly sensitive to the weather anomalies.

If such a transition would occur in an area where the weather is sensitive to the state of solar activity/space weather, and the supplies of agricultural products from outside are for one or another reason limited, the cause-and-effect relation "space weather–prices" described in the present paper can be realized even in the modern conditions of technological progress and market globalization. To identify the regions where this negative process can take place in the near future, progress must be made in understanding the physics of the global climate change and its implications for the local weather conditions in different regions, in our knowledge of the stability thresholds of the area-specific agricultural crops, as well as the further advance in uncovering the mechanisms of solar-terrestrial relationships in the context of their effects on the weather conditions of the studied regions. This task requires combined efforts of experts from various fields: from farmers and meteorologists to astrophysicists and space weather experts.

## ACKNOWLEDGMENTS

The authors thank the Israel Space Agency of the Israel Ministry of Science and Technology for the financial support.